\documentclass{elsart}
\usepackage{natbib}
\usepackage{graphicx}

\begin{document}
\runauthor{Kimura and Bonasera}
\begin{frontmatter}
\title{Bound Electron Screening Corrections to Reactions in Hydrogen Burning Processes} 
\author[infn-lns]{Sachie Kimura}
and 
\author[infn-lns,enna]{Aldo Bonasera}

\address[infn-lns]{Laboratori Nazionali del Sud, INFN,
via Santa Sofia, 62, 95123 Catania, Italy}
\address[enna]{Libera Universit\`{a} Kore di Enna, 94100 Enna, Italy} 

\begin{abstract} 
How important would be a precise assessment of the electron screening effect, on determining
the bare astrophysical $S$-factor~($S_b(E)$) from experimental data? 
We compare the $S_b(E)$ obtained using different screening potentials, 
(1) in the adiabatic limit, (2) without screening corrections, and (3) larger than the adiabatic 
screening potential in the PP-chain reactions.   
We employ two kinds of fitting procedures: 
the first is by the conventional polynomial expression and the second includes explicitly the 
contribution of the nuclear interaction and based on a statistical model.    

Comparing bare $S$-factors that are obtained by using different screening potentials, 
all $S_b(E)$ are found to be in accord within the standard errors for most of reactions investigated,
as long as the same fitting procedure is employed.   
$S_b(E)$ is, practically, insensitive to the magnitude of the screening potential.

\end{abstract}
\begin{keyword}
Astrophysical $S$-factor; Electron screening in the laboratories
\end{keyword}
\end{frontmatter}

\section{Introduction}
On the main-sequence stellar sites the series of reactions that convert hydrogen into helium is known as 
the proton-proton chains. It is a key to understand the evolution of the stars. 
The cross sections of these charged particle induced reactions are major ingredients to calculate 
thermonuclear reaction rates.
They are measured at laboratory energies and are then extrapolated 
to thermal energies~\cite{nacre}, because of their smallness at such low energies. 
The extrapolation is performed by introducing the astrophysical $S$-factor:
\begin{equation}
  \label{eq:sf}
  S(E)=\sigma(E)E e^{2\pi \eta(E)},
\end{equation}
where $\sigma(E)$ is the reaction cross section at the incident center-of-mass (c.m.) energy $E$ 
and $\eta(E)=Z_T Z_P \alpha \sqrt{\frac{\mu c^2}{2E}}$, with $Z_T$, $Z_P$, and $\mu$ denoting the 
atomic numbers and the reduced mass of the target and the projectile~\cite{clayton}; $\alpha$ and $c$ are 
the fine-structure constant and the speed of light, respectively.   
The exponential term in the equation represents the inverse of the Coulomb barrier penetrability.  
Since we have factored out the strong energy dependence of $\sigma(E)$ due to the barrier 
penetrability, the $S$-factor can be approximated by a smooth polynomial expansion in the absence 
of low-energy resonances.
In laboratory experiments, the targets are usually in gas or solid state. 
In the low-energy region, the $S$-factors obtained from experiments show
large enhancement to the extrapolation from high energy data for various reactions~\cite{frvr}.
This enhancement is, usually, attributed to the screening by the bound electrons around the target.
In contrast in the stellar nucleosynthesis nuclei are almost fully ionized and 
surrounded by the plasma electrons. In deuteron induced reactions 
on deuterated metals~\cite{kasagi,raiola1,raiola2} and proton-induced reactions on lithium isotopes 
in several forms of lithium chemical compounds~\cite{cruz}, much larger screening enhancements have 
been observed with respect to the enhancement by gaseous targets.      
The nuclear reactions in such a circumstance are affected by a different 
mechanism of the plasma or the conduction electron screening~\cite{shav,ichimaru,kato:014615}.         
A similar effect has been discussed in the radioactive decay of a nucleus in a model~\cite{PhysRevLett.74.2824}. 
The screening effects in the medium can be dependent on temperature and density of the medium 
and we do not consider such effects in this paper.  
Hence the screening effect of the bound electrons should be removed from the $S$-factor data
to asses the reaction rate in the stellar site correctly.
The enhancement by the bound electrons 
is discussed in terms of a constant potential shift~(screening potential $U_e$).  
The upper limit of $U_e$ is obtained, when the adiabatic approximation is fulfilled,
and it is given by the difference 
of the binding energies of the target atom and the united atom.~\cite{alr}   
On this issue, dynamical effects have been studied 
by following the time evolution of the atomic wave function in the 
classical allowed region~\cite{skls}. 
They solved the time dependent Hartree-Fock equation and evaluated the screening potential.    
Their results suggest that the screening potential 
approaches the adiabatic limit as the incident energy becomes lower. 
The influence of the tunneling phenomenon to this problem has been studied as well~\cite{ktab}. 
And, there, the screening potential could go over the above-mentioned adiabatic limit slightly, 
only in the case where the electronic wave-function has some excited state components at the classical 
turning point of the inter-nuclear motion.   
 We have examined the problem using molecular dynamics approach with constraints~\cite{pmb,kb-ags}, 
to see the effect of the fluctuations~\cite{kb-cdf,kb-icfe}.  
The obtained average enhancement factors do not exceed the adiabatic 
limit. However, there are events that give enhancement factors larger than that in the adiabatic limit.

In this paper, we discuss 
the influence of the electron screening effects on the determination 
of the bare $S$-factors~($S_b(E)$).
We especially focus on the reactions in the hydrogen burning process and 
determine  $S_b(E)$ 
through a fitting of experimental data making use of the polynomial expression and fixing the screening 
enhancement in the adiabatic limit. 
There are several more sophisticated theoretical models that describe the bare $S$-factor
as the $R$-matrix model~\cite{ad,barker,daacv}, the potential model~\cite{PhysRevC.61.025801}, 
and the distorted wave Born approximation.   
However our aim is 
to clarify the effect of the electron shielding on the experimental data of the $S$-factors
rather than to determine the $S$-factor based on a particular theoretical nuclear model.
For this purpose we choose the simplest way: the polynomial expression and try to determine 
$S_b(E)$ in a model independent way.   
We will 
discuss the sensitivity of the $S_b(0)$ values on the choice of the degree of the polynomial and the fitting 
range for each particular reaction. 
We stress that the use of simple polynomial expressions does not carry physical meaning and hence the derived
$S_b(0)$ values should be taken with caution. To make up this deficit, 
we propose another expression in place of the polynomial expression. 
The new expression includes an explicit 
contribution of the nuclear interaction. Moreover it is based on 
a two-step process with 
a compound nucleus~(CN) formation and a statistical choice of the exit channel~\cite{weiss,bon87,bb,kb-up}.
Provided that the energy regions of the astrophysical interest is low, we assume that the $l=0$ partial-wave
component is dominant for most of the reactions. 
To our knowledge, the statistical model calculation has been used to estimate the $S$-factor of the 
radiative proton-capture reactions on Sr isotopes~\cite{PhysRevC.64.065803}. They have used the Hauser-Feshbach 
statistical model code and have compared the theoretical results with the experimental data. 
Because in the Hauser-Feshbach statistical model code, a global level density, which is based on some models, 
is used, one can obtain the absolute value of the $S$-factor. They have found discrepancies between the theoretical 
results and the experimental data, especially in the reactions with target isotopes near the neutron 
shell-closure. In their paper they did not aim to fit their experimental data.       

Electron screening effects have been studied on some specific reactions in the chain,
especially on transfer reactions $^3$He($^3$He,2$p$)$^4$He and $^7$Li($p,\alpha$)$^4$He, which are studied 
including extremely low-energy region~\cite{ju98,erag,erag-b}.  
However there is no systematical study for all of the reactions including the radiative capture reactions. 
We aim to deduce these effects from such 
well studied reactions and estimate the screening  quantitatively on all the reactions in the chain. 
This motivates us to employ the adiabatic limit, rather than leaving the screening potential as a 
fitting parameter as it is customary treated in a series of studies~\cite{ju98,barker}. 
For comparison we also perform the fitting procedures without enhancement factor and by treating the screening potential 
as a parameter.
The obtained $S_b$ at zero incident energy are compared with the results in the NACRE compilation~\cite{nacre}, 
in which the authors employed the screening potential higher than the adiabatic limit, and  
with the results in the $R$-matrix analyses~\cite{daacv}.

This paper is organized as follows,  
In Sec.~\ref{sec:efad} we describe the enhancement factor by the bound electrons within the adiabatic limit briefly.
We list up the reactions in PP-chains in Sec.~\ref{sec:ppchain} and explain how we incorporate the enhancement factor
into the fitting procedure. We, especially, give a detailed account on the second fitting procedure based on 
the statistical model.        
Some reactions are analyzed in this section.
We summarize the paper in Sec.~\ref{sec:sum}.    

\section{Enhancement factor in the adiabatic limit}
\label{sec:efad}
To discuss the enhancement quantitatively,   
we determine the enhancement factor~\cite{alr}:
\begin{equation}
  \label{eq:enh}
  f_e=\frac{\sigma(E)}{\sigma_0(E)},
\end{equation}
in terms of the measured cross section $\sigma(E)$ and the bare cross section $\sigma_0(E)$. 
If one assumes that the effect of the electron screening can be represented 
by the constant shift $U_e$(screening potential) of the potential barrier, the enhancement factor 
is approximated by~\cite{alr,skls},
\begin{equation}
  \label{eq:scpot}
  U_e \sim \frac{E}{\pi\eta(E)} \log \ {f_e}. 
\end{equation}
The $U_e$ can be estimated easily in two limiting cases. 
In one case the inter-nuclear velocity is much higher than that of electrons 
velocity; this limit is called the sudden limit. Within this limit the electron wave function is frozen 
during the reaction. 
In the opposite case where the inter-nuclear motion is much slower than electrons motion,
the bound electrons follow the motion of the nuclei adiabatically. Within this adiabatic limit 
the screening potential is expressed by the difference of the binding energies between 
the initial target atom~($BE_{T}$), and the united atom~($BE_{UA}$), which is formed during the reaction.
\begin{equation}
  \label{eq:uad}
  U_e^{(AD)}=BE_{T}-BE_{UA}
\end{equation}
The screening potential within this limit gives the theoretical upper limit.  However, we should stress that these cases deal with
the ideal situation of a ion impinging on an isolated atom.  This should be the ideal situation in nuclear physics experiments where
very thin targets are used together with a well collimated mono-energetic beam.  However, this situation is not fulfilled if the beam energy is very 
low (which is the case of interest in the present paper) or if the atoms are embedded in a medium such as a metal at a given density and temperature.
In the latter case the nuclear process is expected to be influenced by the rearrangement of the electrons in the metal which will give rise to some
peculiarities similar to the studied radioactive decay in a medium~\cite{PhysRevLett.74.2824}.

\section{Bare $S$-factors of PP-chain reactions}
\label{sec:ppchain}
A list of reactions in PP-chains is shown in Table~\ref{tab:a}. 
\begin{table*}
\caption{Reactions in PP-chains, their minimum incident energy in the c.m. system measured so far
and the enhancement factor within the adiabatic limit at the minimum energy.}
\label{tab:a}
\begin{center}
\begin{tabular}{lrr}
\hline
 reactions & $E_{min}$ (keV) & $f_e^{(AD)}(E_{min})$   \\
\hline
H($p,\beta^+ \nu_e$)D &   &  \\
D($p,\gamma$)$^3$He & 2.52 & 1.07\\
$^3$He($^3$He,2$p$)$^4$He &  20.76 & 1.22 \\[3pt]
$^3$He($\alpha$,$\gamma$)$^7$Be & 93.$^{\#}$, 127.$^\ast$ & 1.02 \\
$^7$Be($e^-$,$\nu_e$)$^7$Li &  &\\
$^7$Li($p,\alpha$)$^4$He & 12.7, 10.$^{\ast\ast}$ & 1.18 \\[3pt]
$^7$Be($p,\gamma$)$^8$B &  115.6 &1.01\\
\hline
\end{tabular}
\end{center}

\vspace*{.6cm}
\noindent
\hspace{2cm} $^{\#}$prompt-$\gamma$ method 
\hspace{0.3cm} $^\ast$activation method 
\hspace{0.3cm} $^{\ast\ast}$THM

\end{table*}
The first reaction H($p,\beta^+ \nu_e$)D involves the $\beta$-decay and has too small cross section to be measured
experimentally. Its $S$-factor is calculated from first principles~\cite{bahcall}. 
We, therefore, concentrate on the other 5 reactions except the electron capture reaction $^7$Be($e^-$,$\nu_e$)$^7$Li. 
In the table the minimum incident energies, measured so-far, for each reaction are also shown. 
For two transfer reactions, $^3$He($^3$He,2$p$)$^4$He and $^7$Li($p,\alpha$)$^4$He,
the cross sections have been measured already including the low-energy region where the screening enhancement 
becomes more than 10\%.  
The other three reactions are radiative-capture reactions, which have even smaller cross sections. 
The $S$-factor of the reaction $^3$He($\alpha$,$\gamma$)$^7$Be has been re-determined with high precision 
recently both by detecting $\gamma$-ray from $^7$Be decay(the activation method)~\cite{bemmerer:122502,gyurky:035805}
and by detecting prompt $\gamma$-ray(the prompt method)~\cite{confortola:065803}.  
Its $S$-factor in the low-energy region is extrapolated from high energy data by the $R$-matrix fitting. 
The reaction $^7$Be($p,\gamma$)$^8$B involves unstable nuclei. The $S$-factor of this reaction has been 
determined by means of the direct capture reaction~\cite{baby:065805,ju03} and the Coulomb dissociation 
method~\cite{PhysRevLett.83.2910}. 
It has been claimed that there is an inconsistency between the results of the two 
methods~\cite{ju03,esbensen:042502}, though the question seems to be resolved by reanalyzing the 
data of the Coulomb dissociation method~\cite{schumann:015806}.                   

If one takes into account modifications due to the nuclear potential and all the contributions from 
partial-waves, the bare $S$-factor can be expressed as~\cite{clayton,kb-up} 
\begin{eqnarray}
  S_b(E)\sim\frac{\pi\hbar^2}{2\mu}\sum_l\Pi_{lf}(E)(2l+1)\exp(W_l), \label{eq:sb0}
\end{eqnarray}
where $\Pi_{lf}(E)$ is the probability to obtain a specified exit channel $f$
from a certain entrance channel $l$ and 
\begin{eqnarray}
  W_l=\frac{4Z_TZ_Pe^2}{\hbar} \sqrt{\frac{\mu}{2E}}
  \left[\sin^{-1}\left(\sqrt{\frac{E}{E_c}}\right) + \sqrt{\frac{E}{E_c}}\sqrt{1-\frac{E}{E_c}} \right] \nonumber \\ 
  - 2\sqrt{\frac{l(l+1)E_l}{E_c}}\left(1-\sqrt{\frac{E}{E_c}}\right), \label{eq:wl}
\end{eqnarray}
with $E_c= Z_TZ_Pe^2/R$ (MeV), $E_l=l(l+1)\hbar^2/(2\mu R^2)$ (MeV fm$^2$), i.e., the heights of the 
Coulomb barrier and the centrifugal potential at the nuclear interaction radius $R$.
We assume an empirical formula $R=d\times R_{N_0}$, where $R_{N_0}=1.4 \times (A_T^{1/3}+A_P^{1/3})$ (fm) and
$d$ is a parameter that takes into account the fact that the nuclear potential has a diffuseness; 
with $A_T$ and $A_P$ denoting the mass numbers of the target and the projectile nuclei. 
We call $d$ radial parameter. 
The exponential term in Eq.~(\ref{eq:sb0}) stands for the penetration factor divided by the pure coulomb penetrability.   
At  zero incident energy limit, Eq.~(\ref{eq:sb0}) reduces to~\cite{kb-up} :
\begin{equation}
 \label{eq:sfN3}
S_b(0)\sim \frac{\pi\hbar^2}{2\mu} e^{\frac{4}{\hbar}\sqrt{2\mu Z_TZ_P e^2 R}}\left[\Pi_{0f}(0) +\sum_{l\ge 1}\Pi_{lf}(0)(2l+1)e^{-\frac{2l(l+1)\hbar}{\sqrt{2\mu Z_TZ_P e^2 R}}}\right] . 
\end{equation}

The conventional polynomial expression for the bare $S$-factor of non-resonant reactions
is associated to the Taylor expansion of Eq.~(\ref{eq:sb0}).
Most reactions in the PP-chain are non-resonant, 
one, therefore, can fit the bare $S$-factor using the enhancement factor: 
\begin{eqnarray}
  S(E)&=&S_b(E)\cdot f_e; \hspace*{1cm}    S_b(E)=S_b(0)+S_1E+S_2E^2+\cdot\cdot\cdot,   \label{eq:sbf}
\end{eqnarray}
in an implementation of the nonlinear least-squares 
algorithm.
For resonant reactions the energy dependence of the $S$-factor is given by the Breit-Wigner formula~\cite{clayton} 
\begin{equation}
  \label{eq:bw}
  S(E)=\frac{\pi\hbar^2}{2\mu}\frac{\omega\Gamma_1(E)\Gamma_2}{(E-E_r)^2+(\Gamma/2)^2}e^{2\pi\eta(E)},
\end{equation}
where 
$\omega=(2J+1)/(2j_1+1)(2j_2+1)$ and $E_r$ are the statistical factor and the resonance energy, 
respectively; $\Gamma_1(E), \Gamma_2$ and $\Gamma$ are the entrance and the exit channels partial widths 
and the total width. The incident energy dependence of $\Gamma_1(E)$ is given again by the penetration factor 
in the case of sub-barrier reactions. One, therefore, can write down 
\begin{equation}
  \label{eq:bw2}
  S(E)\sim\frac{\pi\hbar^2}{2\mu}\frac{c_r e^{W_l}\Pi_{lf}(E)}{(E-E_r)^2+(\Gamma/2)^2},
\end{equation}
where $c_r=\omega\Gamma_2 \frac{3\hbar}{R}\sqrt{\frac{2E_c}{\mu}}$~(MeV$^2$), but we determine $c_r$ from the fitting procedure.
In the case where there are more than one data sets  
we weight the data depending on its standard error.    
As one can easily imagine, higher order terms are important to fit the experimental 
data in the high incident energy region. In this paper    
we limit the fitting energy range less than 1 MeV and 
choose the degree of the polynomial to obtain convergence of the $S_b(0)$ value within the statistical errors. 
We will discuss the sensitivity of $S_b(0)$ on the degree of the polynomial and the fitting 
range for each particular reaction.      

Alternatively, the experimental data are fitted using Eq.~(\ref{eq:sb0}) directly,
instead of the polynomial expression.  For this purpose we derive the incident-energy dependence 
of $\Pi_{lf}(E)$ in Eq.~(\ref{eq:sb0}).
According to Weisskopf model~\cite{weiss,bon87,kb-up}, the probability to obtain a certain exit 
channel $f$ after the CN formation is proportional to 
\begin{equation}
  \label{eq:pif}
  \pi_{f} \propto g_f (T_{k_f}^2+2m_fT_{k_f})\exp\left(-\frac{T_{k_f}\sqrt{a}}{\sqrt{Q_{\rm{CN}}}}\right) \sigma_f^{abs},
\end{equation}
where $g_f, T_{k_f}$, and $m_f$ are the number of states for the spin, the kinetic energy, and the mass, respectively, 
of the lightest reaction product in the exit channel; 
$Q_{\rm{CN}}, \sigma_f^{abs}$, and $a$ are the $Q$-value for the CN formation, 
the cross section of the inverse process, and 
the level density parameter $a\sim A_{\rm{CN}}/8.0$ (MeV$^{-1}$)\cite{weiss}; with $A_{\rm{CN}}$ denoting the mass number of 
the CN.
The absolute value of $\Pi_{lf}(E)$ is, in principle, given by 
\begin{equation}
  \label{eq:pifab}
  \Pi_{lf}(E)=\frac{\pi_f}{\sum_f \pi_{f}},
\end{equation}
where the sum in the denominator is taken over all possible exit channels.  
We, however, avoid calculating the sum and determine $\Pi_{lf}(0)$ from fitting procedures of the experimental data. 
Instead, we scale the incident energy dependence of $\Pi_{lf}(E)$   
\begin{equation}
  \label{eq:pi}
  \Pi_{lf}(E)=\Pi_{lf}(0)\frac{T_{k_f}(E)^2+2m_iT_{k_f}(E)}{T_{k_f}(0)^2+2m_iT_{k_f}(0)}\exp\left(\frac{T_{k_f}(0)\sqrt{a}}{\sqrt{Q_{\rm{CN}}}}-\frac{T_{k_f}(E)\sqrt{a}}{\sqrt{E+Q_{\rm{CN}}}}\right),
\end{equation}
where $T_{k_f}(E)=\frac{A_{\rm{CN}}-A_f}{A_{\rm{CN}}}(E+Q)$, with $Q$, and $A_f$ denoting the reaction $Q$-value, and 
the mass number of the lightest reaction product.
Eq.~(\ref{eq:wl}) is obviously valid only at the incident energy lower than the Coulomb barrier. 
In the fitting procedure using Eq.~(\ref{eq:sb0}), 
$\Pi_{lf}(0)$~($l=0,1,2,\cdots$) and the radial parameter $d$ are treated 
as fitting parameters. 
We assume that the nuclear interaction radius can differ for each partial-wave:
$R_l=d_l\times R_{N_0}$. 
The fitting procedures are performed 
only in the energy region below the barrier. The sum of the partial-waves is taken up to the order with which 
the fit converges, mostly $l=0, 1, 2$ are sufficient. 
We anticipate that the description of the reaction mechanism through the CN formation works well,
in particular, in the reactions with the $l=$0 partial-wave in the entrance channel.~\cite{kb-up}

\subsection{$^3$\rm{He}($^3$\rm{He},2$p$)$^4$\rm{He}}
\label{sec:3he3he}
\begin{figure}
  \includegraphics[height=.7\textheight]{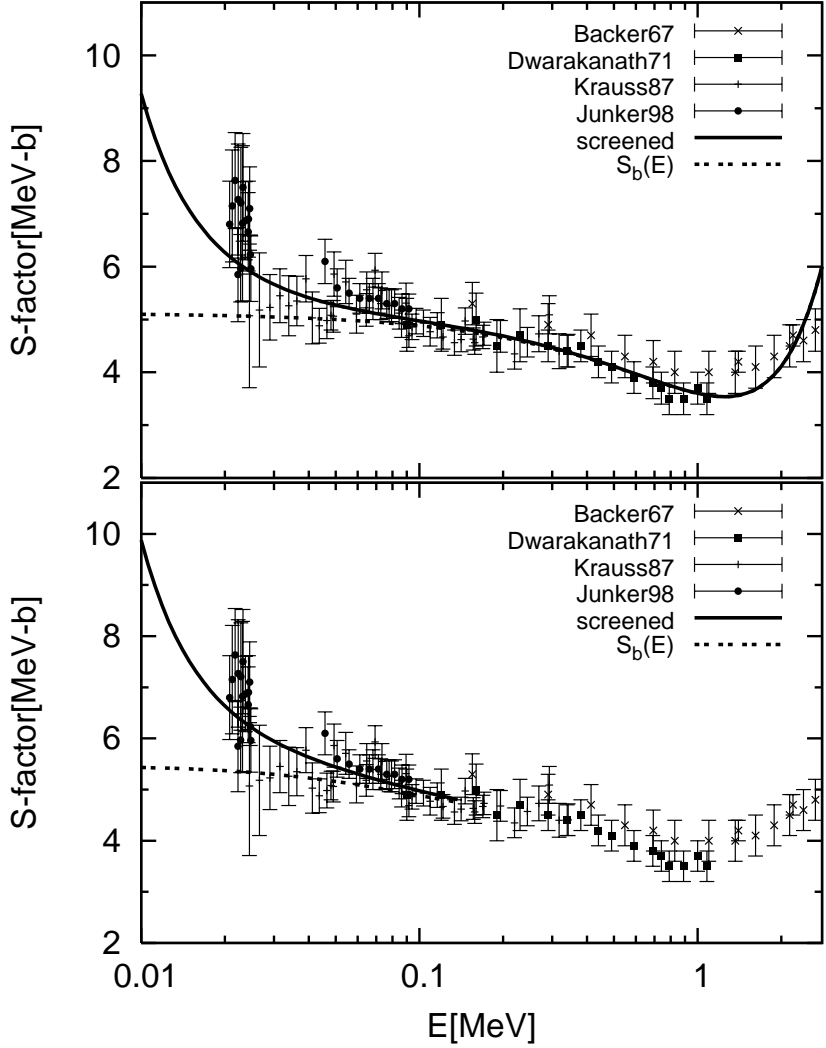}
  \caption{The $S$-factor for the reaction $^3$He($^3$He,2$p$)$^4$He as a function of 
    the incident C.M. energy. The experimental points are 
    from~\cite{ba67}(Backer67), from~\cite{dw71}(Dwarakanath71), from~\cite{kr87}(Krauss87) and 
    from~\cite{ju98}(Junker98).
    The solid curves represent the $S$-factors multiplied by the enhancement factor within the adiabatic limit 
    from our fitting. The dashed curves show the corresponding bare $S$-factor.
    In the top and bottom panels the fitting is performed using the polynomial expression and using 
    Eq.(\ref{eq:sb0}), respectively.}
  \label{fig:3he3he}
\end{figure}

The $S$-factor of the reaction $^3$He($^3$He,2$p$)$^4$He from several measurements 
are shown with error bars in Fig.~\ref{fig:3he3he}. 
At the minimum incident energy, which has been reached in an experiment by the LUNA collaboration~\cite{ju98},
the screening enhancement is estimated to be more than 20\% of the adiabatic approximation.
In the  NACRE~\cite{nacre}compilation, the authors used the screening potential $U_e=$ 330~(eV) 
and a quadratic polynomial 
to obtain the $S$-factor. The fitting parameters in~\cite{nacre} are shown in the first row of table 2.        

For the reaction $^3$He($^3$He,2$p$)$^4$He, the adiabatic screening potential is obtained
under the following considerations. 
In the target medium $^3$He projectiles are likely to be $^3$He$^+$ or $^3$He charge neutral state. 
If we consider $^3$He neutral projectiles, the adiabatic screening potential is 
$U_e^{(AD)}=$ 246.8~(eV)~\cite{ju98}.
\begin{equation}
  f_e^{(AD)}(^3He)=e^{\pi\eta(E)\frac{U_e^{(AD)}}{E}}.
\end{equation} 
For $^3$He$^+$ projectiles the adiabatic screening potential is calculated
taking into account the charge symmetry of the system~\cite{lichten}.       
$U_e^{(AD)1}=$ 255.5~(eV) and $U_e^{(AD)2}=$ 122.2~(eV) in the cases where the system ends up with 
$^6$Be$^+$(1s)$^2$(2s) state and $^6$Be$^+$(1s)(2p)$^2$ state respectively.
The corresponding enhancement factor within the adiabatic limit is written as~\cite{ktab} 
\begin{equation}
  \label{eq:fad2}
  f_e^{(AD)}(^3He^+)=\frac{1}{2}\left(\exp\left[\pi\eta(E)\frac{U_e^{(AD)1}}{E} \right]
                             +\exp\left[\pi\eta(E)\frac{U_e^{(AD)2}}{E} \right]\right).
\end{equation} 
The results of the polynomial fitting using a quadratic polynomial are shown in Table~\ref{tab:3he3he} 
together with fitting parameters in~\cite{nacre}. 
The obtained fitting parameters from the fit without enhancement factor are shown in the 4th row. 
The parameters in the second row are for $^3$He neutral projectile and ones in the third row 
are for $^3$He$^+$ projectile.     
The corresponding curve for $^3$He neutral projectile case is shown in the top panel of Fig.~\ref{fig:3he3he} 
together with the experimental points.  
Notice that the adiabatic limit gives a smaller $S_b(0)$ but  within the standard 
error of the one obtained by the NACRE collaboration.
If we fit the same data by varying the screening potential $U_e$, we obtain $U_e=371 \pm 46$~(eV) 
and $S_b(0)=$ 5.06$\pm$0.09~(MeVb) , with $\chi^2_{\nu}=0.7$. The deduced $U_e$ is higher than the adiabatic limit 
but the obtained $S_b(0)$ is very close to the one in the adiabatic limit.   
Fixing the fitting range from the lowest experimental data 0.0208~MeV to 1~MeV, 
we obtained $S_b(0)$=5.32 $\pm$0.08 (MeVb) using a cubic polynomial with $\chi_{\nu}^2$=0.7.
This $S_b(0)$ coincides with the result using a quadratic polynomial~(the second row in Tab.~\ref{tab:3he3he}).    
Limiting the fitting range from 0.0208~(MeV) to 1~(MeV), $S_b(0)$ is insensitive to the choice of the 
degree of the polynomial. 
Assuming the quadratic polynomial and the adiabatic enhancement factor, 
the value $S_b(0)$ varies from 5.23$\pm$0.06~(MeVb) to 5.26$\pm$0.3~(MeVb), as one changes the 
upper-limit of the fitting from 1~(MeV) to 0.1~(MeV) but fixing the lower limit 0.0208~(MeV). 
On the other hand $S_b(0)$ varies from 5.23~(MeVb) to 5.17~(MeVb), as one changes the lower limit from 0.02~(MeV) to 0.2~(MeV).   
Thus the $S_b(0)$ obtained by using the polynomial expression is not much sensitive to the choice of both the lower and the upper limit.       
\begin{table}
\caption{Fitting parameters of the reaction $^3$He($^3$He,2$p$)$^4$He using polynomial expression.
  The first row is from~\cite{nacre}. The second and the third rows are obtained by using the screening 
  potential in the adiabatic limit. The last row is obtained by assuming without enhancement.}
\label{tab:3he3he}
\begin{tabular}{llllrl}
\hline
  & $S_{b}(0)$(MeVb) & $S_{1}$(b) & $S_{2}$(MeV$^{-1}$b) & $U_e$(eV) & $\chi^2_{\nu}$ \\
\hline
  & 5.18  & -2.22  &  0.80 & 330 & \\
 $f_e^{(AD)}(^3$He) & 5.23 $\pm$ 0.06 & -3.1 $\pm$ 0.5 & 1.6 $\pm$ 0.5 & 246.8 & 0.7 \\
 $f_e^{(AD)}(^3$He$^+)$ & 5.32 $\pm$ 0.06 & -3.5 $\pm$ 0.5 & 1.9 $\pm$ 0.6 & & 0.8 \\
  & 5.56 $\pm$ 0.07 & -4.7 $\pm$ 0.6 &  3.0 $\pm$ 0.7  & 0  & 1.1  \\
\hline         

\end{tabular}
\end{table}

\begin{table}
\caption{Fitting parameters and the $S_{b}(0)$ of the reaction $^3$He($^3$He,2$p$)$^4$He using Eq.~(\ref{eq:sb0}).
  The first row is obtained by using the screening potential in the adiabatic limit, 
  the second row without enhancement, and the last row is obtained by treating $U_e$ as a fitting parameter.}
\label{tab:3he3he-2}
\begin{tabular}{lclcllc}
\hline
   $\Pi_{0f}(0)$ & $\Pi_{1f}(0)$ & $d_0$ & $d_1$ & $U_e$(eV) & $\chi_{\nu}^2$ & $S_{b}(0)$(MeVb) \\
\hline
   0.19 & 7.4$\times 10^{-8}$ & 0.61 & 9.9 & 246.8 & 0.68 & 5.4   \\
   0.17 & 2.2$\times 10^{-8}$  & 0.63 &13.  & 0     & 0.73 & 6.5   \\
   0.18  & 1.0$\times 10^{-7}$ & 0.60 & 9.1 & 299$\pm$102   & 0.68  & 5.2  \\

\hline        
\end{tabular}

\end{table}




The fitting procedures using Eq.~(\ref{eq:sb0}) with the screening potentials $U_e=$~246.8, 0~(eV) 
and treating $U_e$ as a fitting parameter
give zero energy $S$-factors 5.4, 6.5, 5.2 (MeVb), respectively, and they are shown in Table~\ref{tab:3he3he-2}. 
We have used the $l=$0 and 1 partial-waves.
The fitting procedure with $U_e=$ 0 gives 
$\chi^2_{\nu}$ slightly larger than the other two cases. 
If we take into account the enhancement factor,   
the radial parameters $d_{0}$ and $d_{1}$ for two cases are about 0.6 and 9., respectively, 
and $\Pi_{1f}(0)$ is much smaller than $\Pi_{0f}(0)$. 
The former implies that the effective radius of the $l=$1 partial-wave component is larger than that of the 
$l=$ 0 component. And the latter implies that the $l=$ 0 component gives a dominant contribution to the 
$S$-factor. The $l=$1 component plays a major role in the higher energy region.          
The resulting $S_b(0)$ are   
in agreement with the extrapolations using quadratic polynomials for the corresponding screening 
potentials.  
The curve obtained by this fitting procedure for the adiabatic enhancement is shown in the bottom
panel in Fig.~\ref{fig:7lip2}. 
If we fit the same data by varying the screening potential $U_e$, we obtain $U_e=299 \pm 102$~(eV) 
and $S_b(0)=$5.2~(MeVb), with $\chi^2_{\nu}=0.68$.  This $\chi^2_{\nu}$ is the same as 
in the case where we used the adiabatic screening potential.     


\subsection{$^7$\rm{Li}($p,\alpha$)$^4$\rm{He}}
\label{sec:7lip}

\begin{figure}
  \includegraphics[height=.80\textheight]{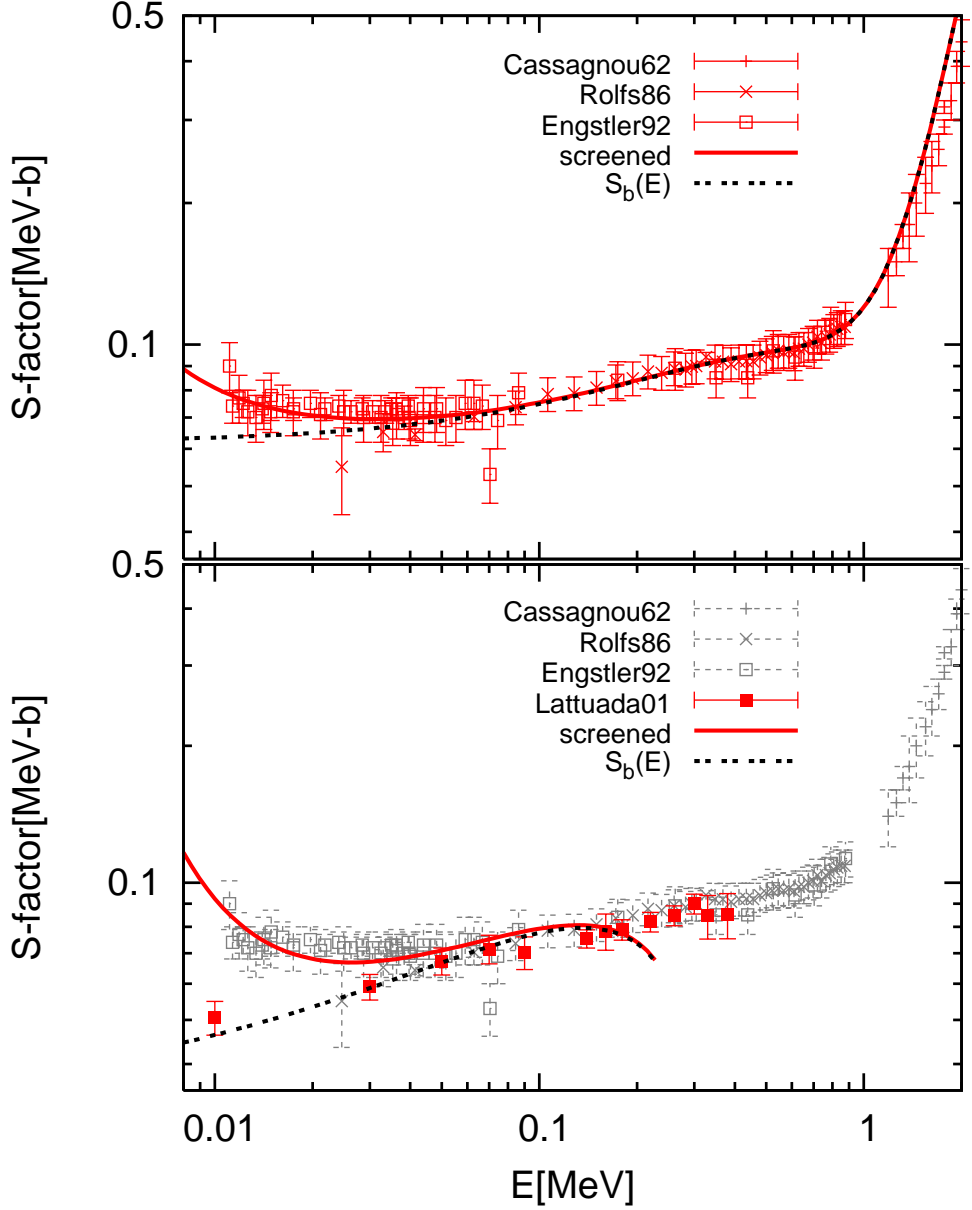}
  \caption{$S$-factor for the reaction $^7$Li($p,\alpha$)$^4$He as a function of the incident c.m. energy. 
    In the top panel experimental points are taken 
    from~\cite{cjmsf}(Cassagnou62), from~\cite{rk86}(Rolfs86), from~\cite{erag,erag-b}(Engstler92). 
    The solid curve represents our fit by polynomial expression with the adiabatic enhancement. 
    The dashed curve corresponds to the bare $S$-factor.    
    In the bottom panel the data from~\cite{la01}(Lattuada01) are shown, together with other experimental 
    data in the top panel.
    The solid curve represents our fit by using Eq.(\ref{eq:sb0}), instead of 
    the polynomial expression, and with treating the screening potential as a parameter.}
  \label{fig:7lip2}
\end{figure}

In Fig.~\ref{fig:7lip2} experimental data of the $S$-factor of the reaction 
$^7$Li($p,\alpha$)$^4$He from several direct measurements 
are shown with error bars. 
We performed the fit of the data 
in the incident energy region from 0.01~MeV to 1~MeV 
using a cubic polynomial without enhancement factor. 
The obtained fitting parameters are shown in the third row of Table~\ref{tab:7liph}. 
This fitting procedure without enhancement factor is quite sensitive to the choice 
of both the upper and the lower limits of the fitting range. 
\begin{table}
\caption{Fitting parameters of the reaction $^7$Li($p,\alpha$)$^4$He. 
  The three rows are obtained by using a cubic polynomial
  and from~\cite{nacre}~(the first row), with the adiabatic approximation(the second row),
  and without enhancement~(the third row).}
\label{tab:7liph}
\begin{tabular}{llllrl}
\hline
  $S_{b}(0)$(MeVb) & $S_{1}$(b) & $S_{2}$(MeV$^{-1}$b) & $S_{3}$(MeV$^{-2}$b) & $U_e$(eV) & $\chi^2_{\nu}$\\
\hline
 0.0593  & 0.193 & -0.355 & 0.236 & 300 & \\
 0.0620 $\pm$ 0.0006 &  0.15 $\pm$ 0.01 & -0.24 $\pm$ 0.03 & 0.14 $\pm$ 0.03 & 175 & 0.41 \\
 0.0673 $\pm$ 0.0008 & 0.10 $\pm$ 0.01 & -0.13 $\pm$ 0.04 & 0.08 $\pm$ 0.03  &  0 & 0.62 \\
\hline
\end{tabular}


\end{table}
\begin{table}
\caption{Fitting parameters and the $S_{b}(0)$ of the reaction $^7$Li($p,\alpha$)$^4$He
  using Eq.~(\ref{eq:sb0}).
  The first row is obtained by using the screening potential 300 eV, the second row in the adiabatic limit, 
  the third row without enhancement, and the last row is obtained by treating $U_e$ as a fitting parameter.}
\label{tab:7lip-2}
\begin{tabular}{llrlc}
\hline
 $\Pi_{1f}(0)$ & $d_{1}$ & $U_e$~(eV)  & $\chi_{\nu}^2$ & $S_{b}(0)$(MeVb) \\
\hline
 7.1$\times$10$^{-5}$ & 4.7  & 300  & 1.4 & 0.033 \\
 6.0$\times$10$^{-5}$ & 4.9  & 175  & 2.1 & 0.035 \\
 5.0$\times$10$^{-5}$ & 5.1  &   0  & 3.5 & 0.038 \\
 9.6$\times$10$^{-5}$ & 4.3  & 495 $\pm$ 41 & 1.0 & 0.031 \\
\hline         
\end{tabular}
\end{table}





The experimental data from~\cite{erag,erag-b} show the enhancement of the $S$-factor in the low-energy region.
In the compilation NACRE~\cite{nacre}, authors used the screening potential $U_e=$ 300~(eV), which is larger than 
the adiabatic limit, and a cubic polynomial to obtain the $S$-factor. The fitting parameters 
in~\cite{nacre} are shown in the first row.       
However in the experiments~\cite{erag,erag-b} LiF solid targets and deuteron projectiles as well as 
deuterium molecular gas targets and Li projectiles are utilized. In the case of LiF target, which is
an ionic crystal similar to NaCl and a large band gap insulator, one can approximate the electronic 
structure of the target $^7$Li state by the $^7$Li$^+$ with 
only two innermost electrons. Thus one expects the screening potential in the adiabatic 
limit $U^{(AD)}_e$ = 371.8-198.2$\sim$174~(eV). 
The best fit of the experimental data in the form of Eq.~(\ref{eq:sbf}) and $U_e^{(AD)}=$ 175~(eV)~\cite{kb-icfe} gives 
the fitting parameters in the second row of Table~\ref{tab:7liph} with the reduced $\chi^2=0.46$. 
The corresponding $S$-factor is shown with the dashed curve in Fig.~\ref{fig:7lip2}.
If we fit the same data by varying the screening potential $U_e$, we obtain $U_e=195\pm 28$~(eV) 
and $S_b(0)=$0.061$\pm$0.001~(MeVb), with $\chi^2_{\nu}=0.41$.  This is, practically, consistent with the result
(the second row of Table~\ref{tab:7liph}) in the adiabatic limit. 
$S_b(0)$ in the adiabatic limit is lower than the value obtained by the fit assuming no screening enhancement 
but higher than the value obtained in~\cite{nacre} where the authors used $U_e=$ 300~(eV).  
We have checked 
the sensitivity of $S_b(0)$ on the degree of the polynomial and the fitting range.
Fixing the fitting range from the lowest experimental data 0.011MeV to 1MeV, 
we obtained $S_b(0)$=0.0620$\pm$0.0006~(MeVb) and 0.0617$\pm$0.0009~(MeVb), using a cubic polynomial 
and using a quartic polynomial, respectively. Both have $\chi_{\nu}^2$=0.41.   
Hence we choose a cubic polynomial for the following fit.  
In addition to this, assuming the adiabatic enhancement factor, 
the value $S_b(0)$ varies from 0.0620~(MeVb) to 0.0632~(MeVb), as one changes the fitting upper limit from 1~(MeV) to 0.5~(MeV)
with fixing the lower limit 0.011~MeV. 
On the other hand $S_b(0)$ varies from 0.0620~(MeVb) to 0.578~(MeVb), as one changes the lower limit from 0.01~(MeV) to 0.04~(MeV).   
$S_b(0)$ obtained by using the polynomial expression is more sensitive to the choice of the lower limit 
than to the choice of the upper limit.

The fitting procedures of the same data but using Eq.~(\ref{eq:sb0}) are performed.  
We have used only the $l=$1 partial-wave, 
because $l$ must be odd to obtain positive-parity state of $^8$Be~\cite{barker2000,daacv}.
Using only the $l=$1 component, our fitting procedure gives a steeper incident 
energy dependence than the case where $l=$0 is used. In passing we mention briefly that the fitting procedure 
of the reaction $^6$Li($p$,$\alpha$)$^3$He in the next subsection for a comparison.
With the screening potential $U_e=$300, 175, 0~(eV)
we obtain zero energy $S$-factors 0.033, 0.035, 0.038~(MeVb), respectively, as they are shown 
in Table~\ref{tab:7lip-2}. They are all considerably smaller than the results of polynomial fitting
and $\chi_{\nu}^2$ of all cases are larger than the results of polynomial fitting. 
The radial parameters $d_{1}$ for all three cases are considerably larger than 1. 
This fact suggests that the interaction radius is 5 times larger than the empirical formula for the 
$l=$1 partial-wave.           
Again, if we fit the same data by varying the screening potential $U_e$, we obtain $U_e=495 \pm 41$~(eV).
The fitting parameters of this procedure are 
shown in the last row in Table~\ref{tab:7lip-2}. 
The curve obtained by this fitting procedure is shown in the bottom
panel in Fig.~\ref{fig:7lip2}. 
The extracted bare $S$-factor data by THM are especially shown with the closed squares~\cite{la01} but 
these data are not included in the fitting procedure. Nevertheless the obtained bare $S$-factor curve follows   
the data by THM, which is thought to give the bare $S$-factor. 

In Ref.~\cite{daacv} the $R$-matrix fitting for higher energy region~($E > 40$~keV) has been used to determine the $S$-factor. 
They obtained the zero-energy $S$-factor 0.067 $\pm$ 0.004~(MeVb) and the screening potential  
$U_e=100\pm25$~(eV), which is less than that within the adiabatic limit.
The $S$-factor at zero energy from our results using polynomials $S_{b}(0)$=0.065 $\pm$ 0.005~(MeVb) 
and 0.0620 $\pm$ 0.0006~(MeVb) are in agreement with this result from the $R$-matrix fitting. 

\subsection{$^6$\rm{Li}($p$,$\alpha$)$^3$\rm{He}}
\label{sec:6lip}

\begin{figure}
  \includegraphics[height=.7\textheight]{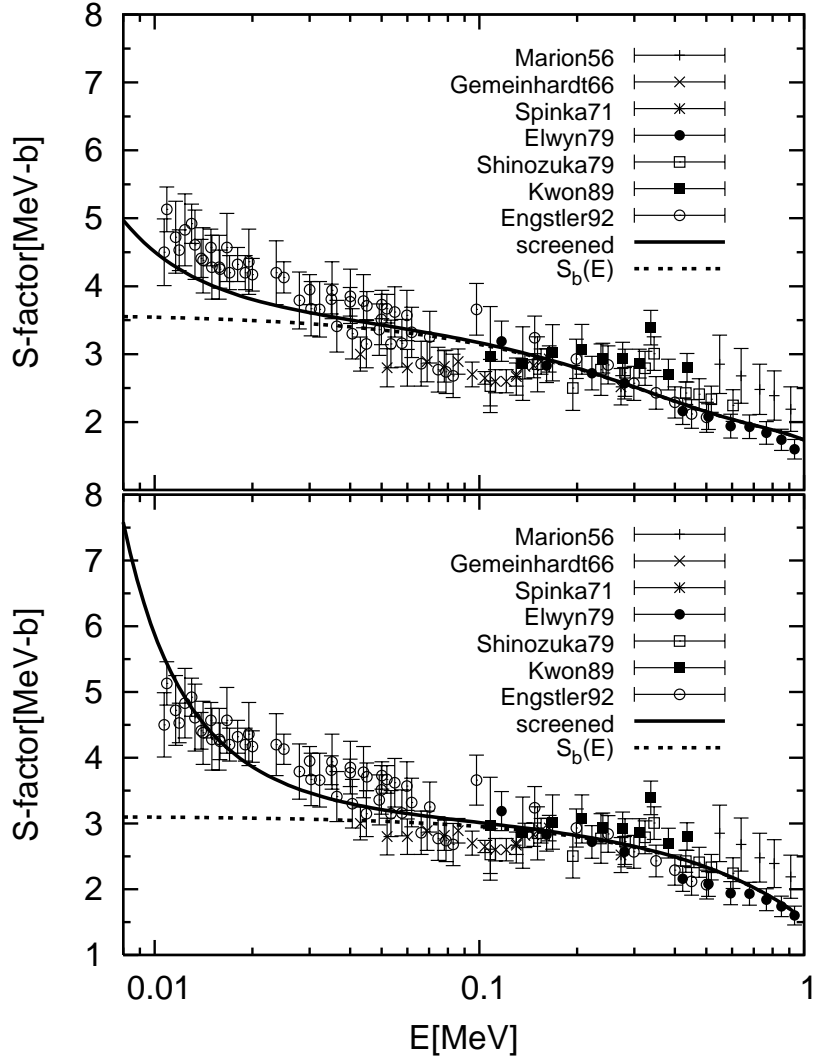}
  \caption{$S$-factor for the reaction $^6$Li($p$,$\alpha$)$^3$He as a function of the incident c.m. energy. 
    The experimental points are from~\cite{ma56}(Marion56), from~\cite{ge66}(Gemeinhardt66), 
    from~\cite{sp71}(Spinka71), from~\cite{el79}(Elwin79), 
    from~\cite{sh79}(Shinozuka79), from~\cite{kw89}(Kwon89) and from~\cite{erag}(Engstler92).
    In the top panel the solid curve represents our fit by polynomial expression with the adiabatic enhancement. 
    The dashed curve corresponds to the bare $S$-factor.
    In the bottom panel the solid curve represents our fit by using Eq.~(\ref{eq:sb0}), instead of 
    the polynomial expression, and with treating the screening potential as a parameter.}
  \label{fig:6lip}
\end{figure}

In contrast to the reaction $^7$Li($p$,$\alpha$)$^4$He, the reaction $^6$Li($p$,$\alpha$)$^3$He does not have 
the restriction on the incident partial-wave. For a comparison, we show the fitting curves of this reaction
in Fig.~\ref{fig:6lip}, where the top panel shows the two curves from a fitting procedure using a polynomial
and the bottom panel shows the curves obtained by using Eq.~(\ref{eq:sb0}), although this 
reaction is not included in the PP-chains, 

The fitting procedures of the experimental data using Eq.~(\ref{eq:sb0}) are performed.  
Using the $l=$0 partial-wave alone, 
we obtain 
$S_b(0)=$~3.3~(MeV), with $\chi^2_{\nu}$ =2.5 with fixing the screening potential 
$U_e$ = 175~(eV) in the adiabatic limit. 
We obtain $U_e$ = 466 $\pm$ 31~(eV) and $S_b(0)=$~3.1~(MeV), with $\chi^2_{\nu}$ =1.4
by treating $U_e$ as a fitting parameter.   
The fitting parameters are shown in Table~\ref{tab:6lip-2}.  
For this reaction the radial parameter $d_{0}$ for all three cases are about unity. 
Although the screening potentials are different, the difference 
does not affect  either the $S_b(0)$ or the fitting parameters, $\Pi_{0f}(0)$ and $d_{0}$. 
Only the $\chi^2_{\nu}$ for $U_e$ = 466 $\pm$ 31~(eV) is much smaller than the others.     
The obtained screening potential from the latter procedure is in agreement with one in the 
reaction $^7$Li($p$,$\alpha$)$^4$He. 
This fact supports the isotopic independence of the electron screening. 

\begin{table}
\caption{Fitting parameters and the $S_{b}(0)$ of the reaction $^6$Li($p,\alpha$)$^3$He by using fitting procedure Eq.~(\ref{eq:sb0}). The first row is obtained by using the screening potential in the adiabatic limit, 
 the second row corresponds to the zero screening potential, 
 the last row is obtained by treating $U_e$ as a fitting parameter.}
\label{tab:6lip-2}
\begin{tabular}{llrlc}
\hline
 $\Pi_{0f}(0)$ & $d_{0}$ & $U_e$~(eV)  & $\chi_{\nu}^2$ & $S_{b}(0)$(MeVb)  \\
\hline
 0.11 & 1.2  & 175  & 2.5 & 3.3 \\
 0.11 & 1.2  &   0  & 3.8 & 3.5 \\
 0.12 & 1.1  & 466 $\pm$ 31 & 1.4 & 3.1 \\
\hline         
\end{tabular}
\end{table}




From the tree results of all the transfer reactions considered, 
$^3$He($^3$He,2$p$)$^4$He, $^7$Li($p,\alpha$)$^4$He, and $^6$Li($p$,$\alpha$)$^3$He, 
one can say that the enhancement by the screening is crucial, in the sense that the fitting 
procedure without enhancement gives $\chi_{\nu}^2$ larger than the others.
However the obtained $S_b(0)$ is insensitive to the magnitude of the screening potential.  

\subsection{\rm{D}($p$,$\gamma$)$^3$\rm{He}}
\label{sec:}
\begin{figure}
  \includegraphics[height=.7\textheight]{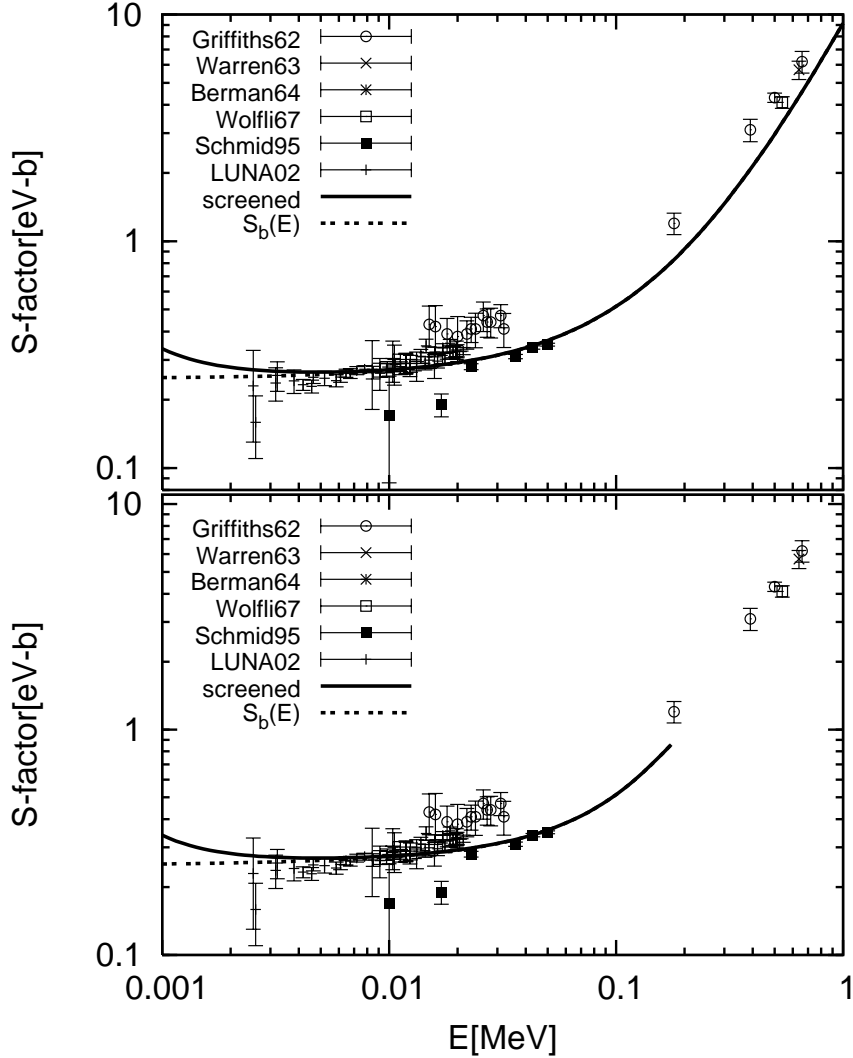}
  \caption{$S$-factor for the reaction D($p$,$\gamma$)$^3$He as a function of the incident c.m. energy. 
    The experimental points are from~\cite{gr62}(Griffiths62), from~\cite{wa63}(Warren63), 
    from~\cite{be64}(Berman64), from~\cite{wo67}(Wolfli67), from~\cite{sc95}(Schmid95) and 
    from~\cite{ca02}(Casella02).
    In the top panel the solid curve represents our fit by polynomial expression with the adiabatic enhancement. 
    The dashed curve corresponds to the bare $S$-factor.
    In the bottom panel the curves are same with ones in the top panel, but using Eq.~(\ref{eq:sb0}) instead of 
    the polynomial expression.}
  \label{fig:dp}
\end{figure}

The $S$-factor data of the reaction D($p$,$\gamma$)$^3$He from several measurements 
are shown with error bars in Fig.~\ref{fig:dp}. 
We performed the fitting procedure of the data 
in the incident energy region from 0.0025~(MeV) to 1~(MeV) 
using a quadratic polynomial without enhancement factor. 
The obtained fitting parameters are shown in the second row of Table~\ref{tab:Dp}. 
In~\cite{nacre} the same polynomial degree has been used but the low-energy data by~\cite{ca02}
were not available at that time. 

At the minimum incident energy in Ref.~\cite{ca02}
the screening enhancement is estimated to be 7\% at utmost. 
This enhancement within the adiabatic limit is, 
again, estimated by using a linear combination of the even and odd states of the electronic wave function, 
reflecting the charge symmetry of the system as Eq.~(\ref{eq:fad2})
where $U_e^{(AD)1}$ and $U_e^{(AD)2}$ are replaced by 40.8~eV and 0.0~eV, respectively.
\begin{table}
\caption{Fitting parameters of the reaction D($p,\gamma$)$^3$He.
  The three rows are obtained by using a quadratic polynomial 
  and from~\cite{nacre}(the first row), without enhancement~(the second row) and with the 
  adiabatic approximation~(the third row).}
\label{tab:Dp}
\begin{tabular}{lllrl}
\hline
  $S_{b}(0)$~(eVb) &$S_{1}$~(b) & $S_{2}$~(eV$^{-1}$b)  & $U_e$~(eV) & $\chi_{\nu}^2$ \\
\hline
 0.20$\pm$0.07 & 5.60$\pm$2.00   & 3.10$\pm$1.10 & 0.0 & (NACRE)  \\
 0.261$\pm$0.006 & 1.3$\pm$0.2 & 12.0$\pm$1.0 & 0.0 & 2.7  \\
 0.256$\pm$0.006 & 1.4$\pm$0.2 & 11.8$\pm$1.0 & 40.8,0.0 & 3.9 \\
\hline         
\end{tabular}
\end{table}
\begin{table}
\caption{Fitting parameters and the $S_{b}(0)$ of the reaction D($p,\gamma$)$^3$He by using the
  fitting procedure Eq.~(\ref{eq:sb0}). 
  The first and the second rows are obtained by the fitting without enhancement, and by using 
  the screening potential $U_e$ in the adiabatic limit, respectively.}
\label{tab:dp-2}
\begin{tabular}{llllrlc}
\hline
   $\Pi_{0f}(0)$ & $\Pi_{1f}(0)$ & $d_0$ & $d_1$ & $U_e$~(eV) & $\chi_{\nu}^2$ & $S_{b}(0)$~(eVb)  \\
\hline
   2.5$\times$10$^{-8}$  & 3.4$\times$10$^{-8}$ & 3.4 & 4.0 &  0.0 & 2.7 & 0.25 \\
   2.3$\times$10$^{-8}$  & 3.6$\times$10$^{-8}$ & 3.7 & 3.9 & 40.8, 0.0 & 2.8 & 0.25 \\
\hline  
\end{tabular}
\end{table}


The fitting parameters obtained using a quadratic polynomial with the adiabatic enhancement are shown in the
third row in Table~\ref{tab:Dp}
and it is shown with the dashed curve in the top panel in Fig.~\ref{fig:dp}. 
Because the enhancement is less than 7\% even at the lowest measured incident energy, 
it changes insignificantly the zero-energy $S$-factor:
$S(0)$=0.261 $\pm$ 0.006~(eVb) obtained by neglecting 
the enhancement differs only slightly from  
the bare $S$-factor at zero-energy $S_{b}(0)$=~0.256 $\pm$ 0.006~(eVb) from our fitting procedure. 
$S_{b}(0)$=~0.256 $\pm$ 0.006~(eVb) is 
slightly higher than
the result from the $R$-matrix fit $S_{b}(0)$=~0.223$\pm$0.010~(eVb) in Ref.~\cite{daacv}, which is obtained 
as a sum of M1 and E1 contributions.
Limiting the fitting range from 0.0025~MeV to 1~MeV, $S_b(0)$ is insensitive to the choice of the 
degree of the polynomial. However 
the $S_b(0)$ obtained using polynomials is quite sensitive to the choice 
of both the upper and the lower limits of the fitting range. 

We performed the fitting procedures using Eq.~(\ref{eq:sb0}).
This fitting procedure without enhancement and with the adiabatic enhancement factor
lead the fitting parameters and the $S_{b}(0)$ in Table~\ref{tab:dp-2}. 
We have used the $l=$~0 and 1 states.
The obtained $S_{b}(0)$ are essentially the same for two cases and are in agreement with the extrapolations 
using quadratic polynomials.  
The radial parameters $d_{0}$ and $d_{1}$ for both cases are larger than one. 
This can be interpreted 
because the effective radius of deuteron is larger than the one given by the empirical formula.    
The curve obtained by this fitting procedure for the adiabatic enhancement is shown in the bottom
panel in Fig.~\ref{fig:dp}. 

\subsection{$^3$\rm{He}($\alpha,\gamma$)$^7$\rm{Be}}
\label{sec:3he4he}
\begin{figure}
  \includegraphics[height=.7\textheight]{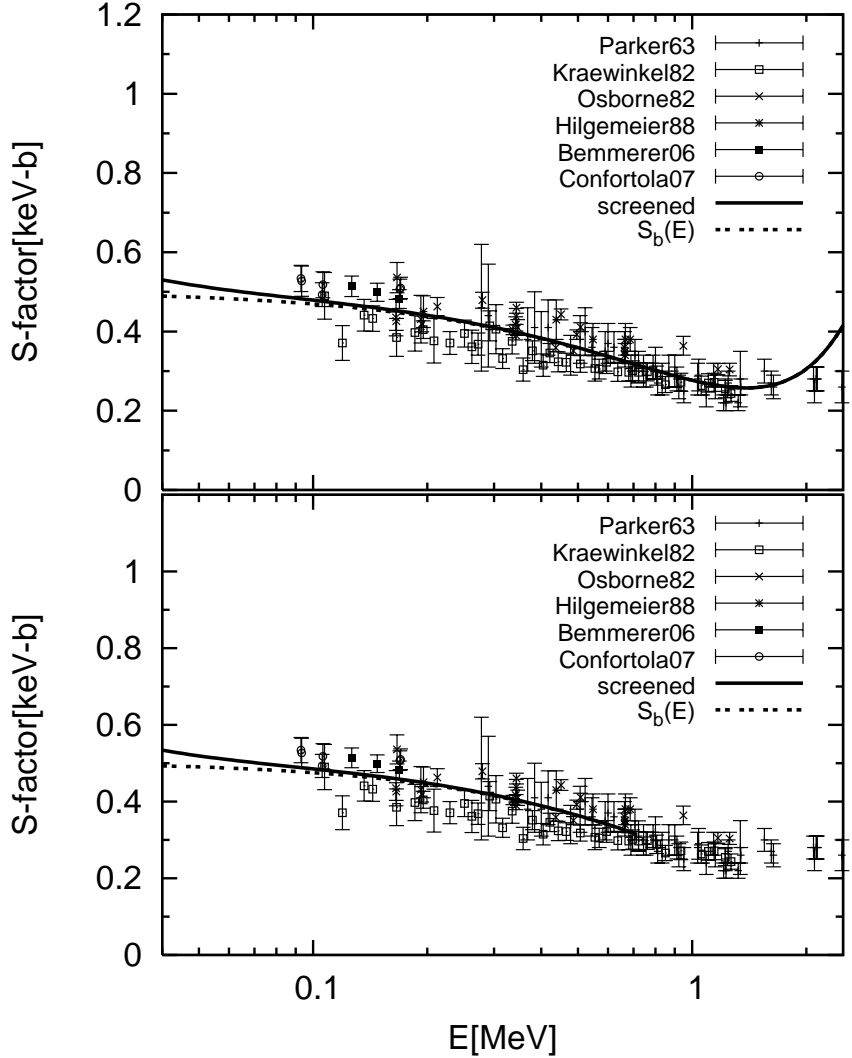}
  \caption{$S$-factor for the reaction $^3$He($\alpha,\gamma$)$^7$Be as a function of the incident c.m. energy. 
    The experimental points are 
    from~\cite{pa63}(Parker63), from~\cite{kr82}(Kraewinkel82), from~\cite{os82}(Osborne82), from~\cite{hi88}(Hilgemeier88),
    from~\cite{bemmerer:122502}(Bemmerer06) and from~\cite{confortola:065803}(Confortola07).
    In the top panel the solid curve represents our fit by polynomial expression with the adiabatic enhancement. 
    The dashed curve corresponds to the bare $S$-factor.
    In the bottom panel the curves are same with ones in the top panel, but using Eq.~(\ref{eq:sb0}) instead of 
    the polynomial expression.}
  \label{fig:3he4he}
\end{figure}

The $S$-factor of the reaction $^3$He($\alpha$,$\gamma$)$^7$Be has been investigated
recently both by the activation~\cite{bemmerer:122502,gyurky:035805} and 
the prompt methods~\cite{confortola:065803}. 
The latter confirmed that there is no discrepancy between the obtained $S$-factors by 
two different methods. 
They have discussed the electron screening enhancement factor in the adiabatic limit 
in~\cite{gyurky:035805}, but the $S$-factor data has not been corrected by the effect.  
The $S$-factor of the reaction $^3$He($\alpha,\gamma$)$^7$Be from several measurements 
are shown with error bars in Fig.~\ref{fig:3he4he}. 
The fitting parameters in~\cite{nacre} are shown in the first row of Table~\ref{tab:3he4he}. 
The screening potential for the reaction $^3$He($\alpha,\gamma$)$^7$Be is estimated in the same way 
with the reaction $^3$He($^3$He,2$p$)$^4$He. 
The estimated enhancement at the minimum incident energy within the adiabatic limit is 2\% at utmost. 
We performed the fit of the data 
in the incident energy region from 0.1072~MeV to 1~MeV 
using quadratic polynomials with the adiabatic enhancement factor. 
The obtained fitting parameters are shown in the second row of Table~\ref{tab:3he4he}. 
We have performed the same fit but without enhancement factor and 
obtained the same $S_b(0)$ as the one with the adiabatic enhancement.  
The obtained $S_b(0)$  coincides with the one in~\cite{nacre} within the error.  
However $S_b(0)$ is rather sensitive to the choice of the fitting range, especially to the choice 
of the lower limit. 
$S_b(0)$ is insensitive to the choice of the degree of polynomial in the selected fitting range. 
In the top panel in Fig.~\ref{fig:3he4he} we have shown the results of the fitting using the adiabatic 
enhancement factor.
The $S$-factor at zero-energy $S_{b}(0)$=0.49 $\pm$ 0.01~(keVb) from our procedure is in agreement with
the result from the $R$-matrix fitting 0.51$\pm$0.04~(keVb) in Ref.~\cite{daacv}. 
\begin{table}
\caption{Fitting parameters of the reaction $^3$He($\alpha,\gamma$)$^7$Be.
  The two rows are obtained by using polynomial expression
  and from~\cite{nacre}~(the first row), with the adiabatic approximation~(the second row).}
\label{tab:3he4he}
\begin{tabular}{lllrl}
\hline
 $S_{b}(0)$ (keVb) & $S_{1}$ (b) & $S_{2}$ (keV$^{-1}$b) & $U_e$ (eV) & $\chi_{\nu}^2$\\
\hline
 0.54 $\pm$ 0.09  & -0.52  & -0.52 &  0.0 & (NACRE)\\
 0.50 $\pm$ 0.01 & -0.35 $\pm$ 0.04 & 0.13 $\pm$ 0.03 & 246.8 & 0.0014\\
\hline         
\end{tabular}
\end{table}


\begin{table}
\caption{Fitting parameters and the $S_{b}(0)$ of the reaction $^3$He($\alpha,\gamma$)$^7$Be 
  by using the fitting procedure Eq.~(\ref{eq:sb0}).
  The first and the second rows are obtained by the fitting without enhancement, and by using 
  the screening potential $U_e$ in the adiabatic limit, respectively.}
\label{tab:3he4he-2}
\begin{tabular}{llllrlc}
\hline
   $\Pi_{0f}(0)$ & $\Pi_{2f}(0)$ & $d_0$  & $d_2 $  & $U_e$~(eV) & $\chi_{\nu}^2$ & $S_{b}(0)$~(keVb)  \\
\hline
   5.1$\times$10$^{-7}$ & 4.9$\times$10$^{-7}$ & 4.0 & 2.5 & 0.0  & 2.3 & 0.51 \\
   5.3$\times$10$^{-7}$ & 4.9$\times$10$^{-7}$ & 4.0 & 2.5 & 246.8  & 2.3 & 0.50 \\
\hline  
\end{tabular}
\end{table}



The fitting parameters from the fitting procedures with Eq.~(\ref{eq:sb0}) are shown 
in Table~\ref{tab:3he4he-2}. We have used the $l=$0, 2 partial-wave contributions. 
This procedure using the enhancement factors with $U_e^{(AD)}=$ 246.8, 0.0~(eV) gives zero 
energy $S$-factors 0.50 and 0.51~(keVb), respectively. Both are in agreement with the result 
from the polynomial fitting procedure.
The radial parameters $d_{0}$ and $d_{2}$ for both cases are larger than one. 
The curve obtained by this fitting procedure with the adiabatic enhancement is shown in the bottom
panel in Fig.~\ref{fig:3he4he}. 
The obtained zero energy $S$-factor is smaller than $S(0)=0.560\pm 0.017$~(keV) in~\cite{confortola:065803}. 
This is because their result is obtained by a normalization to their data.  

\subsection{$^7$\rm{Be}($p$,$\gamma$)$^8$\rm{B}}
\label{sec:7bep}
The reaction $^7$Be($p$,$\gamma$)$^8$B is a key process to produce the high energy solar neutrinos through 
the $\beta$-decay of $^8$B. The $S$-factor of this reaction is investigated intensively by many groups by means 
of the direct capture~(DC) 
reaction~\cite{baby:065805,ju03}, the indirect Coulomb dissociation~(CD) method~\cite{PhysRevLett.83.2910,schumann:015806}
and the asymptotic normalization coefficients~(ANCs)~\cite{trache:062801}.  
It was claimed that the experimental data of the $S$-factor by CD experiments gives a steeper energy dependence than that by 
DC experiments in the low-energy region and the lower zero-energy $S$-factor as an average~\cite{ju03}. 
Recently this inconsistency has been resolved by reanalyzing data by the Coulomb dissociation method~\cite{schumann:015806}.

\begin{figure}
  \includegraphics[height=.7\textheight]{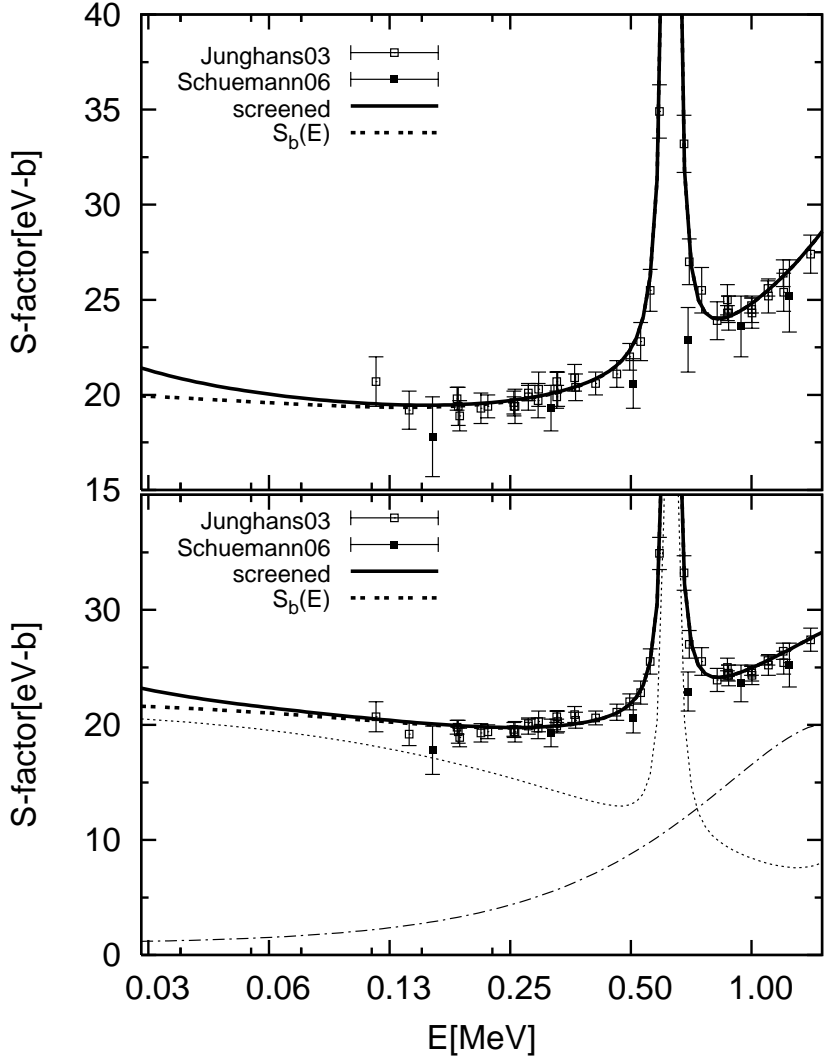}
  \caption{$S$-factor for the reaction $^7$Be($p$,$\gamma$)$^8$B as a function of the incident c.m. energy. 
    The experimental points are from~\cite{ju03}(Junghans03) and from~\cite{schumann:015806}(Schuemann06).
    In the top panel the solid curve represents our fit by polynomial expression with the adiabatic enhancement. 
    The dashed curve corresponds to the bare $S$-factor.
    In the bottom panel the curves are same with ones in the top panel, but using Eq.~(\ref{eq:sb0}) instead of 
    the polynomial expression.}
  \label{fig:7bep}
\end{figure}
For the purpose of the extrapolation of zero-energy $S$-factor,  
both the DC and the CD experiments above mentioned use the microscopic cluster model~\cite{desc} and
give the zero-energy $S$-factor 22.1 $\pm$ 0.6(expt.) $\pm$ 0.6(theor.)~(eVb)
[$21.4 \pm 0.5$(expt.) $\pm$ 0.6(theor.)~(eVb) as a mean of all modern direct measurements] 
and 20.6 $\pm 0.8$~(stat.) $\pm$ 1.2~(syst.) (eVb), respectively. 
Let us remind you that the purpose of this paper is not a precise 
determination of the $S$-factor but to see the effect of the electron screening on the 
determination of the $S$-factor.  
So that we rather use the consistent approaches with the other reactions than employ a special 
treatment for this reaction. 
However we, at least, need to include the resonances to analyzing the DC data~\cite{ju03}.
For this purpose we use the Breit-Wigner formula Eq.~(\ref{eq:bw}).   
Moreover it is well known that this reaction has a low-energy bound state in the $^8$B~\cite{PhysRevC.58.3711, PhysRevC.61.045801}. To take into account this state, we use 
\begin{eqnarray}
  S_b(E)= \frac{S_{-1}}{E_B+E}+S_0+S_1E,   \label{eq:sbf2w}
\end{eqnarray}
where $E_B=0.1375$ (MeV)~\cite{PhysRevC.58.3711}, in place of the polynomial expression. 
We make a special mention of $^7$Be metallic target being used in the experiment~\cite{ju03}.    
The experimental data of the $S$-factor in Ref.~\cite{ju03} is fitted with 
the Breit-Wigner single-level resonance formula for 1$^+$ and 3$^+$ resonances plus Eq.~(\ref{eq:sbf2w}). 
We use the resonance parameters in Ref.~\cite{ju03} and 
obtained $S_b(0)=$ 20.8~(eVb) with $\chi_{\nu}^2=$0.3 from the fitting procedure 
without enhancement factor. 
Assuming the adiabatic enhancement,
the screening enhancement factor is of the order of 1\% at the minimum incident energy of the 
experiment in~\cite{ju03}. By making use of the enhancement factor with the adiabatic screening 
potential $U_e^{(AD)}=$ 222.0~(eV), we obtain $S_b(0)=$ 20.5~(eVb) with $\chi_{\nu}^2=$0.3.
The corresponding bare $S$-factor is shown with the dashed curve in the top panel in Fig.~\ref{fig:7bep}.
The difference between two $S_b(0)$ using different screening potentials is less than 2\%.
Considering the isotopic independence of the electron screening,
we, tentatively, use the screening potential obtained by the measurement of the reaction 
$^9$Be($p,\alpha$)$^6$Li: $U_e$=900$\pm$50~(eV)~\cite{zahnow}.
The fitting procedure gives $S_b(0)=$ 19.7~(eVb) with $\chi_{\nu}^2=$ 0.3.
This is 4\% smaller than the former two results. 
We summarize the fitting parameters in above fitting procedures in Table~\ref{tab:7bep}.

\begin{table}
\caption{Fitting parameters and $S_b(0)$ of the reaction $^7$Be($p$,$\gamma$)$^8$B by using 
  the polynomial Eq.~(\ref{eq:sbf2w}).  The four rows are obtained from~\cite{nacre}(the first row), 
  without enhancement~(the second row), with the adiabatic approximation~(the third row),
  and with the screening potential $U_e$= 900~(eV)~\cite{zahnow}~(the fourth row).} 
\begin{tabular}{llllrll}
\hline
 {$S_{-1}$~(eV$^2$b) }  & {$S_0$~(eVb) } & {$S_{1}$~(b) } & {$S_{2}$ } & $U_e$~(eV) & $\chi^2_{\nu}$ & $S_b(0)$\\
\hline
 &  21. $\pm$ 2. & 18. & 38. & 0.0  & (NACRE) & 21. $\pm$ 2.  \\
 0.6 $\pm$ 0.2 & 15.7 $\pm$ 0.6 & 8.3 $\pm$ 0.4 & & 0.0 & 0.3 & 20.8 \\
 0.7 $\pm$ 0.2 & 15.7 $\pm$ 0.6 & 8.3 $\pm$ 0.4 & & 222.0 & 0.3 & 20.5  \\
 0.7 $\pm$ 0.2 & 16.0 $\pm$ 0.6 & 8.1 $\pm$ 0.4 & & 900.0 & 0.3 & 19.7  \\
\hline
\end{tabular}
\label{tab:7bep}
\end{table}




\begin{table}
\caption{Fitting parameters and the $S_{b}(0)$ of the reaction $^7$Be($p,\gamma$)$^8$B by using the
  fitting procedure Eq.~(\ref{eq:sb0}) with resonant terms. 
  The three rows are obtained by the fitting without enhancement~(the first), and by using 
  the screening potential $U_e$ in the adiabatic limit~(the second), and 
  with the screening potential $U_e$= 900~(eV)~(the third), respectively.}
\label{tab:7bep-2}
\begin{tabular}{llllrlc}
\hline
   $\Pi_{0f}(0)$ & $\Pi_{2f}(0)$ & $d_0$  & $d_2$ & $U_e$~(eV) & $\chi_{\nu}^2$ & $S_0$~(eVb)  \\
\hline
   2.2$\times$10$^{-8}$ & 2.5$\times$10$^{-4}$ & 0.9 & 0.2 & 0.0 & 0.4 & 22.4 \\
   2.2$\times$10$^{-8}$ & 2.3$\times$10$^{-4}$ & 0.9 & 0.2 & 222. & 0.4 & 22.3 \\
   2.2$\times$10$^{-8}$ & 2.0$\times$10$^{-4}$ & 0.9 & 0.2 & 900. & 0.5 & 22.1 \\
\hline  
\end{tabular}
\end{table}

We perform the fitting procedures of the experimental data in Ref.~\cite{ju03} 
using Eq.~(\ref{eq:sb0}) including 1$^+$ and 3$^+$ 
resonances plus another resonance with negative energy -0.1375~(MeV), which corresponds 
to the pole term in Eq.~(\ref{eq:sbf2w}). 
\begin{eqnarray}
  \label{eq:fres}
  S_b(E)&=&\frac{\pi\hbar^2}{2\mu}(\Pi_{0f}(E)e^{W_0}+5\Pi_{2f}(E)e^{W_2})  \nonumber \\
  &+&\frac{c_{r1}e^{W_0}\Pi_{0f}(E)}{(E+0.1375)^2} 
  +\frac{c_{r2}e^{W_0}\Pi_{0f}(E)}{(E-0.630)^2+\Gamma^2/4}
  +\frac{c_{r3}e^{W_2}\Pi_{2f}(E)}{(E-2.183)^2+\Gamma^2/4}
\end{eqnarray}
where $c_{r1}, c_{r2}$ and $c_{r3}$ are scaling factors in Eq.~(\ref{eq:bw2}).
Our fitting procedures using different screening potentials give common 
$c_{r1}=5.\times 10^{-7}$ (MeV$^2$), $c_{r2}=2.\times 10^{-9} $(MeV$^2$) and 
$c_{r1}=2.\times 10^{-6}$ (MeV$^2$).  
In the bottom panel of Fig.~\ref{fig:7bep} 
the thin dotted and dot-dashed curves show the contributions from resonances, including 
the negative-energy resonance, and the non-resonant part, respectively.       
We have used the $l=$~0 and 2 partial-waves.

The fitting procedure using Eq.~(\ref{eq:fres}) 
without enhancement, with the adiabatic enhancement factor, and with $U_e$= 900~(eV) 
lead the fitting parameters and the $S_{b}(0)$ in Table~\ref{tab:7bep-2}. 
Despite of the difference of the utilized screening potentials, neither the obtained zero-energy 
$S$-factor nor $\chi_{\nu}^2$ does have much differences.  
The values $\Pi_{2f}(0)$ are much larger than $\Pi_{0f}(0)$, i.e., 
the $d$-wave contribution is dominant in the energy region investigated.  
The radial parameter $d_{0}$ is smaller than one, and $d_{1}$ is smaller than $d_{0}$ for 
all three cases. 
The curve obtained by this fitting procedure for the adiabatic enhancement is shown in the bottom
panel in Fig.~\ref{fig:7bep}.
The obtained $S_b(0)$ using different screening potentials are in accordance with the result with
the microscopic cluster model~\cite{desc} $S_b$(0)=22.1 $\pm$ 0.6(expt.) $\pm$ 0.6(theor.)~(eVb) in Ref.~\cite{ju03} within the error-bar.

From the tree results of all the radiative capture reactions considered, 
D($p$,$\gamma$)$^3$He, $^3$He($\alpha,\gamma$)$^7$Be, and $^7$Be($p,\gamma$)$^8$B, 
one can say that the obtained $S_b(0)$ is insensitive if 
the screening enhancement, within the adiabatic approximation, is taken into account or not. 
For all the reactions investigated in this paper the fitting procedure with the polynomial 
expression is more sensitive to the difference of the screening potential than the fitting procedure 
with Eq.~(\ref{eq:sb0}).

\section{Conclusions}
\label{sec:sum}
We discussed the bound electron screening corrections to the bare $S$-factors of the reactions in PP-chains.
Our approach is based on fitting procedures of the experimental data.    
For this purpose we employed two different fitting procedures: one is the conventional polynomial expressions
and the other includes explicitly the contribution of the nuclear interaction and based on the statistical model
to describe exit channels. 
The later fitting procedure works especially well for the reactions that have a dominant $s$-wave entrance channel component. 
We have applied different types of screening enhancements: 
in the adiabatic limit, determined through a fit and larger than adiabatic 
screening potentials, as well. 
From the tree results of all the transfer reactions considered, 
$^3$He($^3$He,2$p$)$^4$He, $^7$Li($p,\alpha$)$^4$He, and $^6$Li($p$,$\alpha$)$^3$He, 
the enhancement by the screening is crucial, in the sense that the fitting 
procedure without enhancement gives $\chi_{\nu}^2$ larger than the others in which the screening 
enhancement is taken into account. However    
the obtained $S_b(0)$ is insensitive to the magnitude of the screening potential.  
Especially for the radiative capture reactions D($p$,$\gamma$)$^3$He
and $^7$Be($p,\gamma$)$^8$B, 
the screening correction within the adiabatic approximation has been considered for the first time. 
However the results  
suggest that the obtained $S_b(0)$ is insensitive whether the screening enhancement, within the adiabatic 
approximation, is taken into account or not. 
Making a comparative study of the bare $S$-factors obtained by two-ways of fitting procedures 
using different screening enhancement factors, we found that 
all $S_b(E)$ coincide within the standard errors.  
$S_b(E)$ is, practically, insensitive to the magnitude of the screening potential.

\bigskip 

The authors acknowledge Prof. S. Kubono for the suggestion of the problem and valuable comments.
One of us~(S. K.) thanks Dr. H. Costantini and Dr. R. G. Pizzone for stimulating discussions and 
for providing us experimental data. 



%



\end{document}